\documentclass[prb,amsmath,amssymb,lengthcheck,showpacs,floatfix,groupedaddress,superscriptaddress,longbibliography]{revtex4-1}
\usepackage{graphicx}
\usepackage[english]{babel}
\usepackage{times}
\usepackage{dcolumn}
\usepackage[usenames,dvipsnames]{color}
\usepackage{units}
\usepackage{bm}

\begin{document}

\title{Kondo effect at low electron density and high particle-hole
asymmetry in 1D, 2D, and 3D}

\author{Rok \v{Z}itko}

\affiliation{Jo\v{z}ef Stefan Institute, Jamova 39, SI-1000 Ljubljana, Slovenia}
\affiliation{Faculty  of Mathematics and Physics, University of Ljubljana,
Jadranska 19, SI-1000 Ljubljana, Slovenia}

\author{Alen Horvat}

\affiliation{Jo\v{z}ef Stefan Institute, Jamova 39, SI-1000 Ljubljana, Slovenia}

\date{\today}

\begin{abstract}
Using the perturbative scaling equations and the numerical
renormalization group, we study the characteristic energy scales in
the Kondo impurity problem as a function of the exchange coupling
constant $J$ and the conduction-band electron density. We discuss the
relation between the energy gain (impurity binding energy) $\Delta E$
and the Kondo temperature $T_K$. We find that the two are proportional
only for large values of $J$, whereas in the weak-coupling limit the
energy gain is quadratic in $J$, while the Kondo temperature is
exponentially small. The exact relation between the two quantities
depends on the detailed form of the density of states of the band. In
the limit of low electron density the Kondo screening is affected by
the strong particle-hole asymmetry due to the presence of the
band-edge van Hove singularities. We consider the cases of 1D, 2D, and
3D tight-binding lattices (linear chain, square lattice, cubic
lattice) with inverse-square-root, step function, and square-root
onsets of the density of states that are characteristic of the
respective dimensionalities. We always find two different regimes
depending on whether $T_K$ is higher or lower than $\mu$, the chemical
potential measured from the bottom of the band. For 2D and 3D, we find
a sigmoidal cross-over between the large-$J$ and small-$J$ asymptotics
in $\Delta E$, and a clear separation between $\Delta E$ and $T_K$ for
$T_K < \mu$. For 1D, there is in addition a sizable intermediate-$J$
regime where the Kondo temperature is quadratic in $J$ due to the
diverging density of states at the band edge. Furthermore, we find
that in 1D the particle-hole asymmetry leads to a large decrease of
$T_K$ compared to the standard result obtained by approximating the
density of states to be constant (flat-band approximation), while in
3D the opposite is the case; this is due to the non-trivial interplay
of the exchange and potential scattering renormalization in the
presence of particle-hole asymmetry. The 2D square lattice DOS behaves
to a very good approximation as a band with constant density of
states.
\end{abstract}


\maketitle

\newcommand{\vc}[1]{{\mathbf{#1}}}
\renewcommand{\Im}{\mathrm{Im}}
\renewcommand{\Re}{\mathrm{Re}}
\newcommand{\corr}[1]{\langle\langle #1 \rangle\rangle}
\newcommand{\expv}[1]{\langle #1 \rangle}
\newcommand{\ket}[1]{| #1 \rangle}
\newcommand{\Tr}{\mathrm{Tr}}

\section{Introduction}

In experimental setups designed for studying the Kondo effect in
quantum dots and other nanostructures, the leads attached to the
device are typically low-dimensional electron systems such as a
two-dimensional electron gas or even an effectively one-dimensional
nanowire. Since the electron density in such systems is tunable, it
may occur that at low enough electron density the chemical potential
lies close to the band-edge singularities in the density of states
(DOS) of the reservoir. In such a situation there is an extreme
asymmetry between the particle and hole states \cite{shchadilova2014}.
This affects the Kondo screening of the local moment and leads to
significant deviations from the standard results obtained in the
flat-band approximation that assumes a featureless (i.e., constant)
DOS \cite{hewson}.

This work explores the Kondo physics at low electron density $n$ so
that the chemical potential $\mu$ is close to the bottom of the band.
We discuss three paradigmatic types of band-edge van Hove
singularities: the inverse square root divergency at the bottom of
one-dimensional (1D) DOS, the step-function singularity in the
two-dimensional (2D) DOS, and the square-root onset in the the
three-dimensional (3D) DOS. These energy-dependencies are directly
dictated by the dimensionality of the system and always occur
(assuming that the dispersion relation $\epsilon(k)$ is differentiable
close to its global minimum in the Brillouin zone, which is generally
the case). To be specific, we use the 1D tight-binding chain DOS, the
2D square-lattice DOS, and the 3D cubic-lattice DOS, while the
magnetic impurity is described using the spin-1/2 Kondo model. The
problem is studied using the numerical renormalization group (NRG)
\cite{wilson1975,krishna1980a,bulla2008} with a discretization
technique that is applicable to arbitrary DOS without any systematic
discretization errors \cite{resolution,odesolv}. We compare numerical
results to the predictions from the perturbative renormalization group
arguments \cite{shchadilova2014}. We pay particular attention to the
different behavior of the energy gain due to the impurity, $\Delta E$,
and the Kondo temperature, $T_K$. $\Delta E$ is calculated by
subtracting the ground state energy of the system with and without the
impurity,
\begin{equation}
\Delta E=E_0(0)-E_0(J).
\end{equation}
This is, hence, the energy required to remove the impurity from the
system (i.e., an impurity binding energy). $T_K$ is determined as the
temperature where the effective impurity moment is reduced to small
numbers, defined according to Wilson as \cite{wilson1975,krishna1980a}
\begin{equation}
T_K \chi_\mathrm{imp}(T_K)=0.07,
\end{equation}
where $\chi_\mathrm{imp}$ is the impurity contribution to the magnetic
susceptibility,
\begin{equation}
\chi_\mathrm{imp} = \expv{S_z^2}_J - \expv{S_z^2}_0
\end{equation}
and $0.07$ is some essentially arbitrary small number. It will be seen
that $\Delta E$ and $T_K$ are in general not proportional to each
other and have different asymptotic behavior in the weak-coupling
limit. We also present a scaling argument that explains the reduction
of $T_K$ in 1D and its enhancement in 3D for small density: they are
due to the {\it curvature} in the DOS close to band edges. In the
appendices we consider the effect of the magnetic field and of the
anisotropy in the Kondo coupling. We also calculate analytically the
quadratic energy gain due to a magnetic impurity in the $J\to0$ limit
in the flat-band approximation.

\section{Model}

We consider the Kondo impurity model
\begin{equation}
H = J \vc{s} \cdot \vc{S} + \sum_{\vc{k}a} \epsilon_\vc{k} c^\dag_{\vc{k}a}
c_{\vc{k}a},
\end{equation}
where $J$ is the Kondo exchange coupling constant,
\begin{equation}
\vc{s}=\frac{1}{2} \sum_{ab} f^\dag_{a} \boldsymbol{\sigma}_{ab} f_{b}
\end{equation}
is the local spin-density at the position of the impurity, with $a,b
\in \{\uparrow,\downarrow\}$ and the local operator at the origin $f$
defined as
\begin{equation}
f^\dag_{a} = \frac{1}{\sqrt{N}} \sum_\vc{k} c^\dag_{ka},
\end{equation}
$N$ being the number of sites forming the lattice, $\vc{S}$ is the
impurity spin-1/2 operator,
\begin{equation}
\vc{S}=\frac{1}{2}\boldsymbol{\sigma},
\end{equation}
and the conduction-band part of the Hamiltonian corresponds to either
a 1D chain, a 2D square-lattice, or a 3D cubic-lattice tight-binding
model. The only information about the conduction band that is relevant
for the impurity model is its density of states $\rho(\epsilon)$,
which is known in analytical form for all three dimensions. The
chemical potential is measured from the bottom of the band, thus
$\mu=0$ corresponds to a completely depleted band. 

The calculations are done within the grand-canonical ensemble in the
true thermodynamic limit, thus both the system size and the total
number of electrons are infinite: the band filling $n$ is controlled
by $\mu$, while $L, N \to \infty$. The NRG discretization is performed
using the method described in Ref.~\onlinecite{resolution} which
correctly handles arbitrary DOS even in the presence of singularities
\cite{vanhove}. This approach allows to calculate the energy gain
$\Delta E$ to extremely high accuracy \cite{groundstate} by performing
two NRG runs, one for a finite $J$ and another for a reference system
with decoupled impurity ($J=0$), then subtracting the obtained
ground-state energies of the discrete Wilson-chain representations;
while these two energies depend significantly on the details of the
NRG discretization, their difference does not and, in fact, changes
very little with varying value of the discretization parameter
$\Lambda$ if the appropriate discretization scheme is used
\cite{groundstate}. The numerical calculations are performed for
$N_z=8$ twisted discretization grids and with the discretization
parameter $\Lambda=2$ \cite{wilson1975,resolution}. The truncation
cutoffs are taken high enough so that the results are fully converged.
The temperature is effectively zero. The results will be presented in
units of half-bandwidth $D=2dt$, where $d$ is dimensionality of the
tight-binding lattice.

\section{Characteristic energy scales}

\subsection{Binding energy $\Delta E$}

\begin{figure}
\centering
\includegraphics[clip,width=0.5\textwidth]{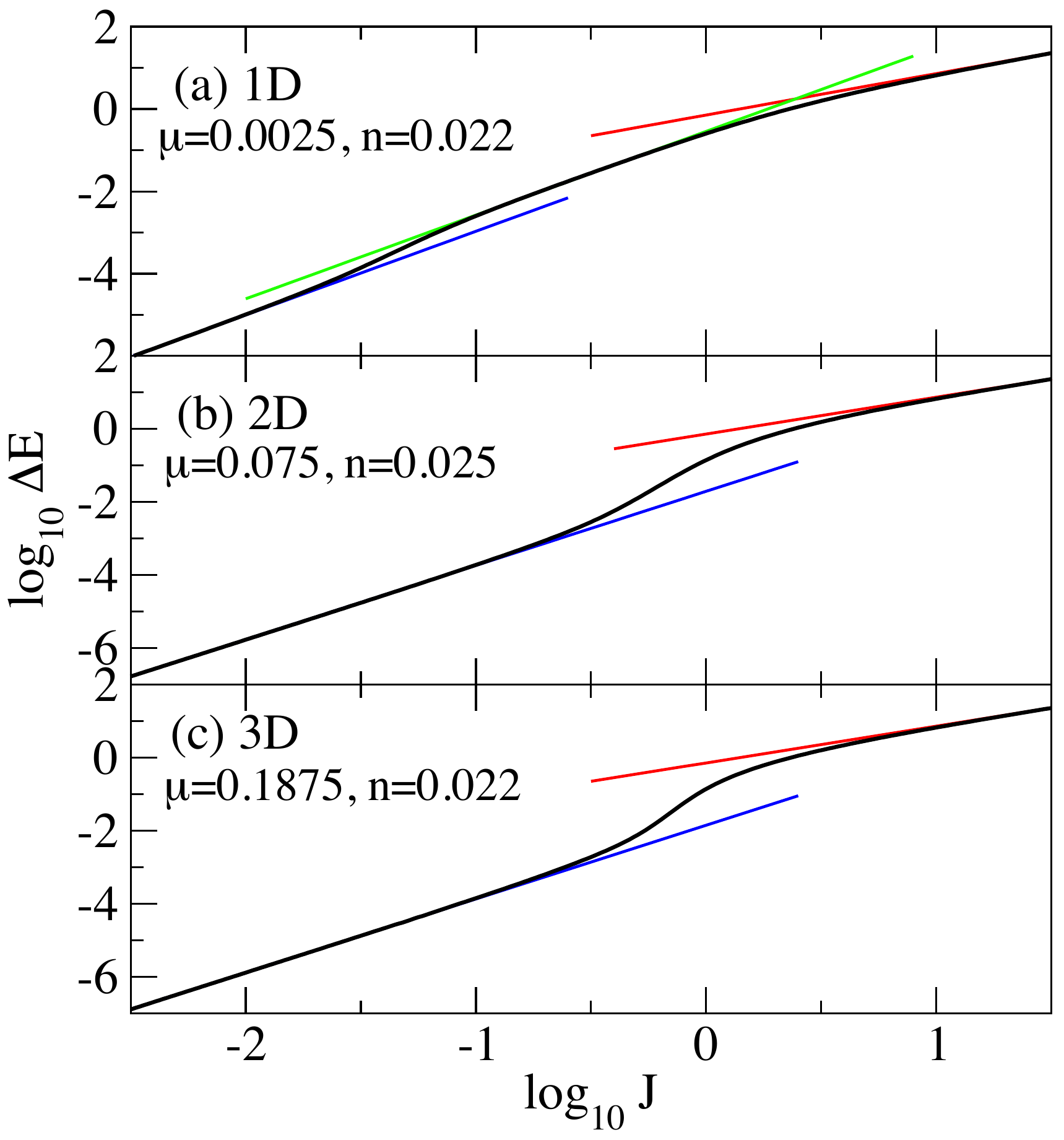}
\caption{Overview of the different parameter regimes in the 1D, 2D,
and 3D cases. The energy gain $\Delta E$ is plotted as a function of
the Kondo exchange coupling $J$ on a log-log scale. The red and blue
lines are the high-$J$ and low-$J$ asymptotics, $\propto J$ and
$\propto J^2$, respectively, common for all three dimensions. For 1D
case in panel (a) we also indicate the intermediate-$J$ regime which
is $\propto J^2$ (green line). The model parameters are indicated in
the respective panels. They are chosen so that the band occupancy is
comparable in all three cases, $n \approx 0.02$.}
\label{fig1}
\end{figure}

We first discuss the relevant parameter regimes of the exchange
coupling $J$. We find that in all cases, irrespective of the
energy-dependence of the DOS and the nature of the singularities, the
small-$J$ and large-$J$ regimes behave exactly the same, see
Fig.~\ref{fig1}.

In the large-$J$ regime, we find the expected result that the energy
gain is $\Delta E = 3J/4$ due to the formation of a local singlet
state between the impurity spin and the lattice site to which the
impurity is attached. This regime is reached for $J \gg D$. In all
cases considered, the numerical value of $\Delta E$ lies atop the
$3J/4$ line within the accuracy of linewidth for $J/D \gtrsim 10$.

In the small-$J$ regime we find quadratic behavior with a
non-universal prefactor $\alpha$ that depends on the DOS
($d=1,2,3$) and the band occupancy $n$,
\begin{equation}
\Delta E = \alpha_d(n) J^2.
\end{equation}
It should be emphasized that $\Delta E$ is not of the order of $T_K$
for small $J$, as often claimed. The difference is perhaps
underappreciated, because the common intuition is that the energy gain
in the formation of the Kondo singlet state should be approximately
$T_K$ since the singlet forms on that energy scale. In fact, the
situation is somewhat more subtle because of the logarithmic scaling
of the exchange coupling in the Kondo mechanism
\cite{kondo1967,hewson}. Many decades of bulk excitations above $T_K$
(in the scaling regime $T_K \ll \epsilon \ll J$) are already slightly
perturbed by the exchange scattering off the magnetic impurity. The
degree of perturbance can be quantified through the quasiparticle
phase-shift of the bath electron with energy $\epsilon$,
$\delta(\epsilon)$. The phase shift for $\epsilon \gg J$ is
essentially zero, because in this range the bath electron spin is
hardly affected by the impurity. The local moment starts to be felt on
the energy scale $\epsilon \sim J$, but the effect is weak and the
phase shift is $\sim J/D$. At still lower energies, the phase shift
then increases logarithmically, until for $\epsilon$ on the scale of
$T_K$ it reaches saturated values of the order of $1$ (recall that
$\delta(\epsilon=0)=\pi/2$ for the Kondo effect at the particle-hole
symmetric point). The total energy gain is given as an integral of the
single-particle energy shifts approximately proportional to
$\delta(\epsilon)$, over all energies $\epsilon$, and since the
contribution is $\sim J$ for all $\epsilon \lesssim J$, we expect the
result to scale as $\Delta E \propto J^2$. The behavior $\Delta E
\propto J^2$ is indeed seen in the numerical calculations of this
quantity using the NRG in all models of this class. Since the Kondo
problem is known to be non-perturbative in $J$, the actual expression
must in fact be of the form $\Delta E = f(J) + g(J)$, where $f(J)$ is
an analytical function which is quadratic in the $J\to0$ limit, while
$g(J)$ is a non-analytical exponentially-small correction which may be
neglected in the $J\to 0$ limit compared to the first term. The
leading correction due to dynamical processes in $f(J)$ is actually
$O(J^3)$, hence analytical, see Appendix B.

\begin{figure}
\centering
\includegraphics[clip,width=0.5\textwidth]{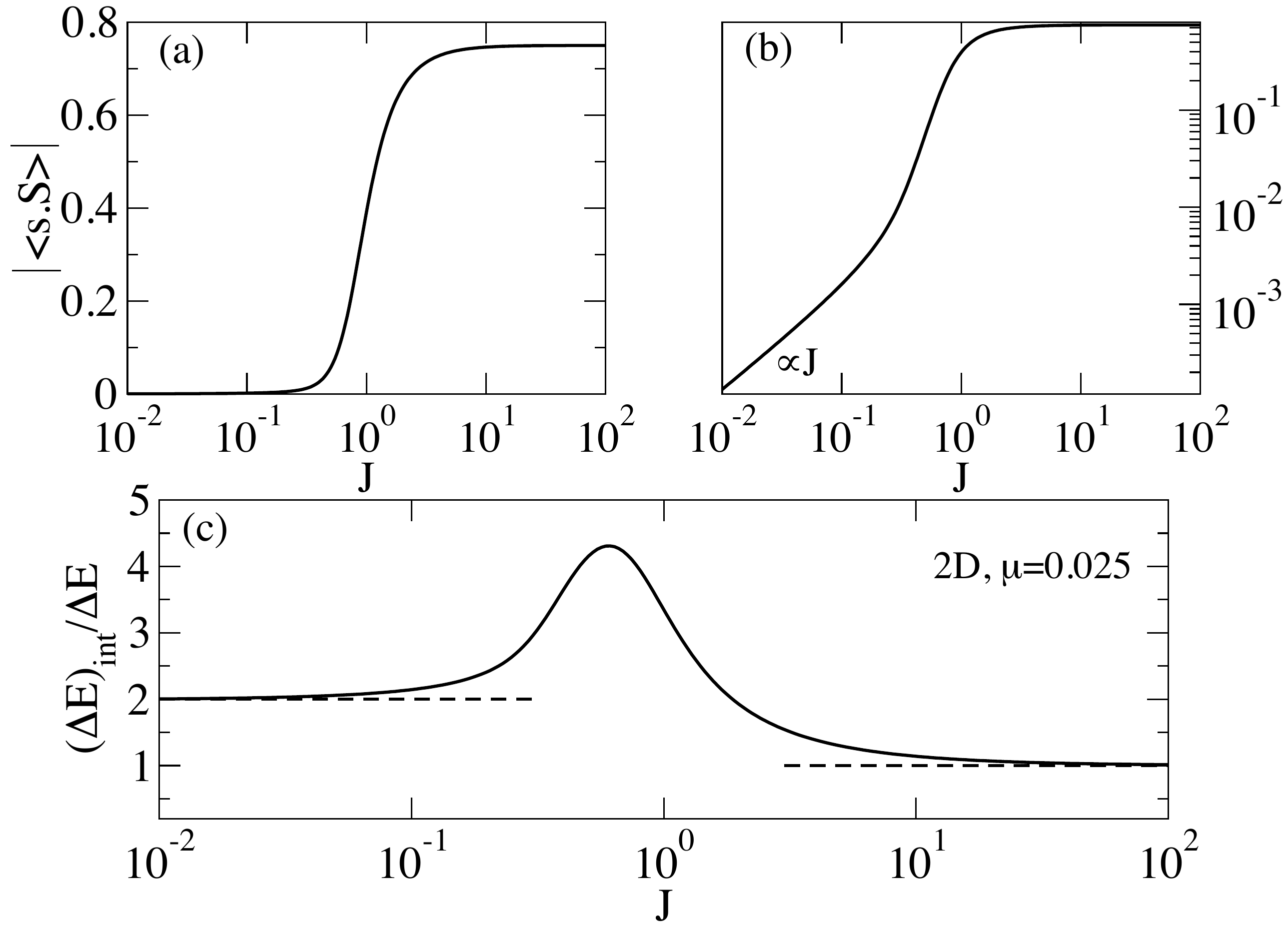}
\caption{(a,b) Absolute value of the spin-spin expectation value,
$|\expv{\vc{s} \cdot \vc{S}}|$, i.e., the correlation between the
impurity spin and the conduction band spin at the location of the
impurity, plotted on log-linear and log-log scales. (c) Ratio between
the local contribution to the energy gain $(\Delta E)_\mathrm{int} = J
|\expv{\vc{s} \cdot \vc{S}}| \propto J^2$, and the total energy gain
$\Delta E$. The calculations are performed for 2D DOS with
$\mu=0.025$.}
\label{misc1}
\end{figure}

Another quantity of interest is the expectation value of $\expv{\vc{S}
\cdot \vc{s}}$, i.e., of the interaction term in the Kondo
Hamiltonian. Since this term is coupled linearly to $J$, it can be
estimated through $d\expv{H}/dJ=-d\Delta E/dJ$. Since $\expv{\vc{S}
\cdot \vc{s}} \propto J$, see Fig.~\ref{misc1}(a,b), by integration of
$\Delta E$ over $J$ we again find $\Delta E \propto J^2$. In fact, it
is interesting to compare $\Delta E$ with the quantity
\begin{equation}
(\Delta E)_\mathrm{int} = J |\expv{\vc{S} \cdot \vc{s}}|,
\end{equation}
see Fig.~\ref{misc1}(c). As expected, in the large-$J$ limit the full
contribution to the energy gain comes from the singlet localized at
the position of the impurity. The low-$J$ limit of the ratio $(\Delta
E)_\mathrm{int}/\Delta E$ is exactly $2$; this is found universally
for all values of filling and for all three dimensionalities. In fact,
this follows directly from the quadratic behavior of $\Delta E$.
Namely, 
\begin{equation}
\frac{(\Delta E)_\mathrm{int}}{\Delta E} = \frac{J (d\Delta
E/dJ)}{\Delta E} = \frac{d\ln\Delta E}{d\ln J},
\end{equation}
thus the plot in Fig.~\ref{misc1}(c) actually represents the
logarithnic derivative of $\Delta E$ from which we can directly read
off the local power-law exponent $\beta$ at given $J$. It then follows
that
\begin{equation}
\label{eqnr}
\begin{split}
(\Delta E)_\mathrm{int} &= \beta \Delta E, \\
(\Delta E)_\mathrm{kin} &= (1-\beta) \Delta E,
\end{split}
\end{equation}
where the kinetic part $(\Delta E)_\mathrm{kin}=(\Delta E)-(\Delta
E)_\mathrm{int}$. This result indicates that asymptotically, for
$J\to0$, twice the binding energy is always contributed by the local
$\vc{s}(\vc{r}=0)$ term, while the kinetic part $(\Delta
E)_\mathrm{kin}$ is actually negative and equal in absolute value to
$\Delta E$.  In the high-$J$ limit, we have $\beta=1$, thus the energy
gain is entirely due to the local interaction term with no correction
from the bulk. Fig.~\ref{misc1}(c) also indicates that $\beta > 1$ at
any finite $J$, thus the binding (energy gain) is always due to the
interaction term, while the kinetic term always reduces the binding
since it is strictly negative.

Returning now to the discussion of the different regimes of $J$ in
relation to Fig.~\ref{fig1}, we note a clear difference between the 2D
and 3D compared to the 1D case. In 2D and 3D, the low-$J$ and high-$J$
asymptotic regimes of $\Delta E$ are connected by a single sigmoidal
cross-over curve whose inflection point is slightly below $J = 1$, and
the low-$J$ asymptotic regime is reached at $J \approx 0.2$. In 1D,
the behavior is qualitatively different: at $J \approx 3$ we observe a
very smooth crossover from the $3J/4$ behavior to an intermediate
$J^2$ regime which extends to very low $J$ values, then at $J \approx
0.03$ there is a crossover to the asymptotic weak-coupling $J^2$
regime with a different prefactor, see Fig.~\ref{fig1}(a). This is, in
fact, the physics discussed in Ref.~\onlinecite{shchadilova2014}: when
the characteristic energy scale, such as $\Delta E$ or $T_K$, is
larger than $\mu$, the system is sensitive to the diverging DOS at the
bottom of the conduction band in one dimension and the system is in a
different universality class, namely the class associated with the
density of states that diverges as a power-law at the Fermi level
\cite{Mitchell:2013jo}. When the characteristic scale is below $\mu$,
however, we recover the more conventional Kondo screening behavior as
found in 2D and 3D.

\subsection{Kondo temperature $T_K$}

\begin{figure}
\centering
\includegraphics[clip,width=0.5\textwidth]{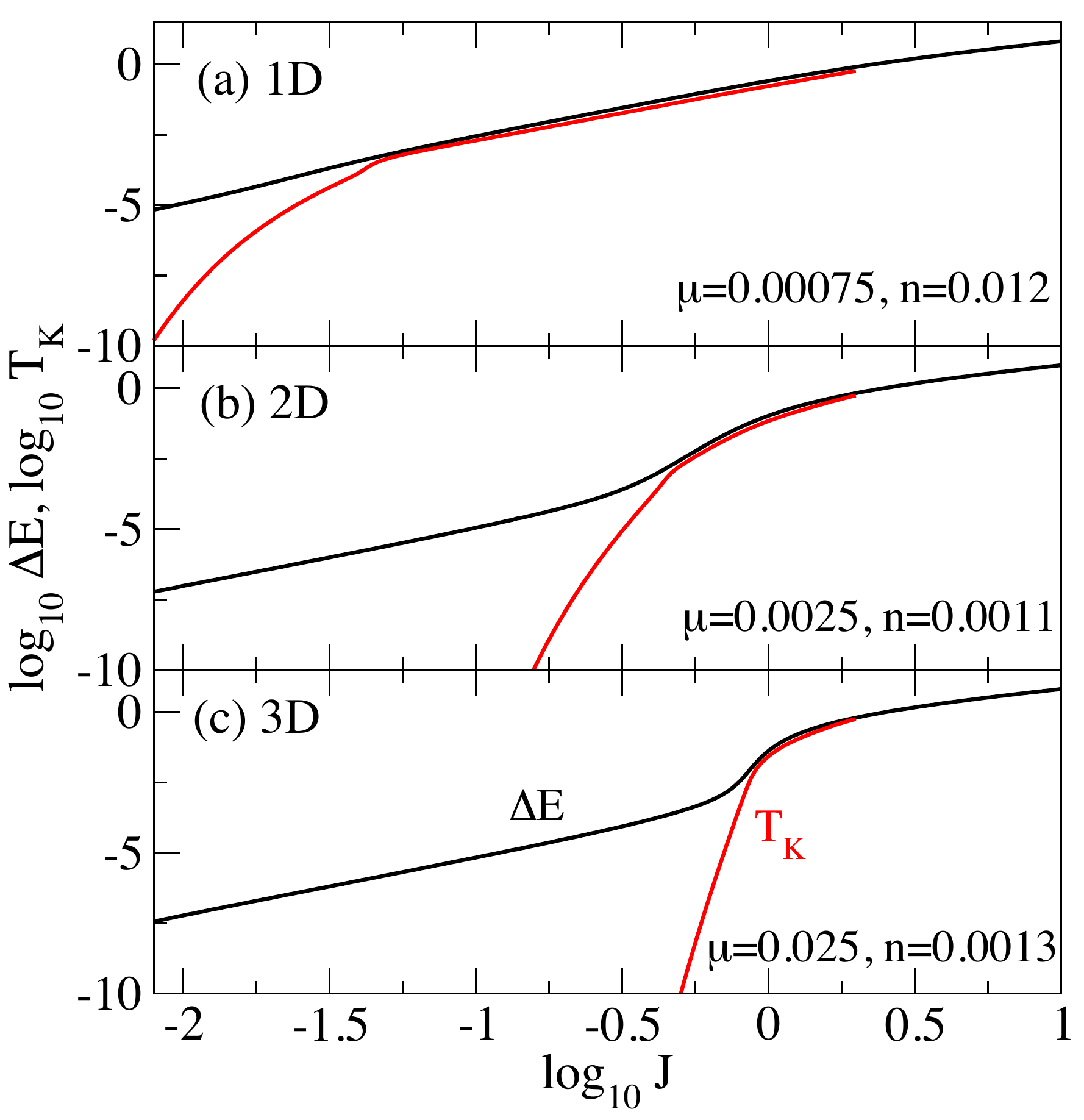}
\caption{Relation between the energy gain $\Delta E$ and the Kondo
temperature $T_K$ for three different dimensionalities. The two
quantities behave similarly only in the intermediate-coupling regime,
but have very different weak-coupling asymptotic behavior.}
\label{fig2}
\end{figure}

We now directly compare the energy gain $\Delta E$ and the Kondo
temperature $T_K$ in Fig.~\ref{fig2}. We recall that $T_K$ is defined
thermodynamically through the impurity magnetic susceptibility
according to Wilson's prescription $T_K \chi_\mathrm{imp}(T_K) = 0.07$
\cite{wilson1975}. This is thus the temperature scale on which the
impurity magnetic moment becomes quenched, irrespective of the
mechanism how this actually occurs (via Kondo effect of conventional
or unconventional type, or via local singlet formation). We find that
the two quantities are proportional in the cross-over intermediate-$J$
and large-$J$ regimes, $\Delta E \approx c T_K$, where $c$ is a
non-universal prefactor of the order of $1$. For large $J$, $T_K$
cannot be directly calculated in the NRG, but our definition through
moment quenching is consistent with $T_K$ behaving as $T_K \sim J$ in
the large-$J$ limit, thus in this sense the proportionality persists
there. For small $J$, however, $T_K$ behaves exponentially,
\begin{equation}
T_K \sim \exp(-1/\rho_0 J),
\end{equation}
while $\Delta E$ is quadratic, as discussed above. (Here
$\rho_0=\rho(\mu)$ is the density of states at the Fermi level, which
depends on the value of electron density $n$.) Indeed, we find that
the lower-$J$ boundary of the intermediate-$J$ regime is defined
precisely by the point where the two quantities run apart. This is
especially pronounced in 1D \cite{shchadilova2014}, but also occurs in
2D and 3D.

\subsection{Onset of the Kondo scaling regime}

\begin{figure}
\centering
\includegraphics[clip,width=0.5\textwidth]{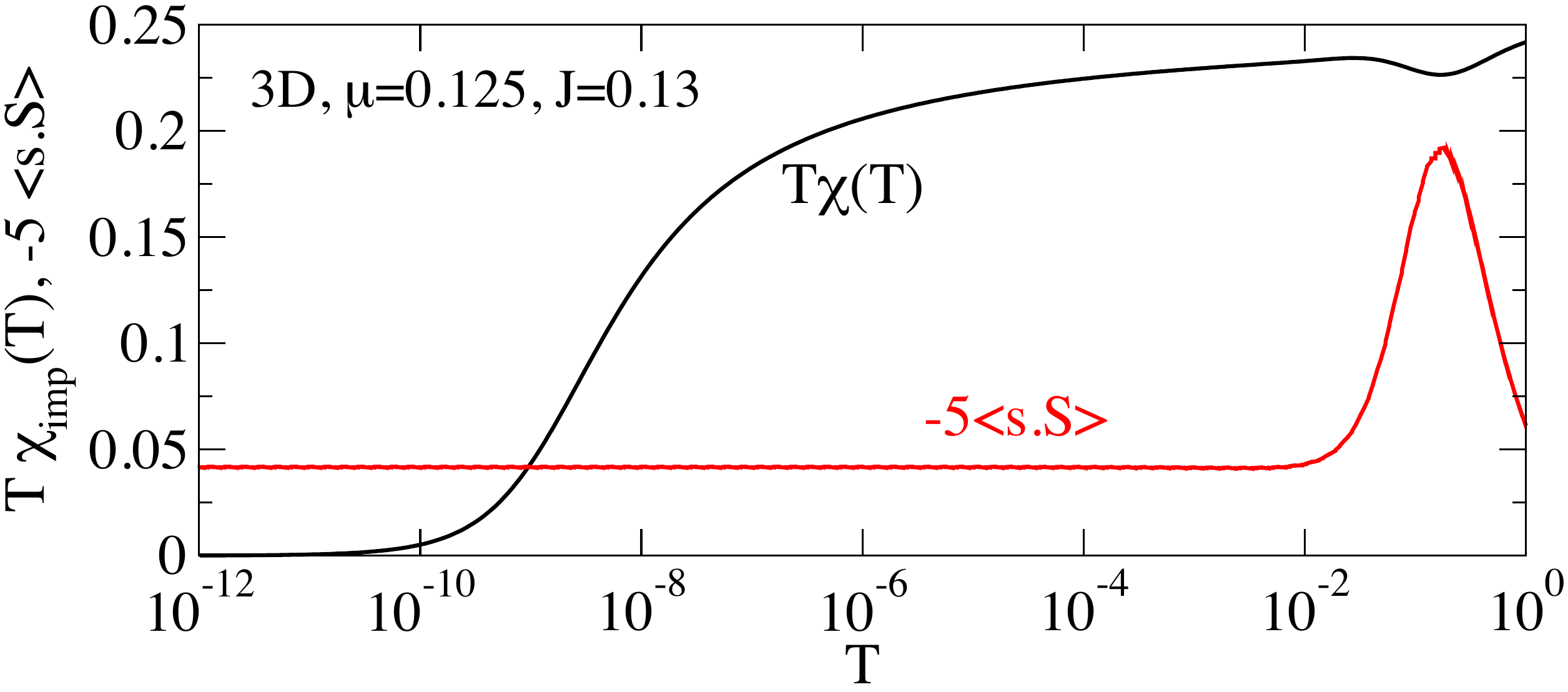}
\caption{Temperature dependence of the impurity magnetic
susceptibility $T\chi_\mathrm{imp}(T)$ and of the spin-spin correlation
$\expv{\vc{s}\cdot\vc{S}}$, in the parameter regime where $\Delta E$
and $T_K$ are already significantly separated.}
\label{misc2}
\end{figure}

The difference between $\Delta E$ and $T_K$ in the weak-coupling limit
for $T_K < \mu$ should not be too surprising, because the Kondo effect
is something which happens in the conduction band due to the presence
of the impurity, possibly involving bulk electrons far away from its
position, while the impurity binding is more locally sensitive. It is
also interesting to compare the temperature dependences of the
impurity susceptibility $T\chi_\mathrm{imp}(T)$ and of the expectation
value $\expv{\vc{s}\cdot\vc{S}}(T)$, see Fig.~\ref{misc2}. It is found
that the expectation value reaches its saturated value on the scale $T
\sim 10^{-2}$, significantly above $\Delta E$ which is here equal to
$4 \times 10^{-4}$, while the impurity local moment is quenched at
much lower temperatures. 

If, however, a similar comparison is performed in the parameter range
where $\Delta E$ and $T_K$ are proportional, i.e., for $T_K > \mu$, we
find that both $T\chi_\mathrm{imp}(T)$ and $\expv{\vc{s} \cdot
\vc{S}}(T)$ saturate on the same temperature scale. In addition, we
find that $T\chi_\mathrm{imp}(T)$ does not show universal Kondo
behavior at low temperatures in such cases. 

From these observations we conclude that the separation between
$\Delta E$ and $T_K$ is actually a sign of the onset of a true
conventional Kondo effect where the spin is quenched through the
gradual scaling of the effective exchange coupling $J$ from a small
bare value $J$ to the strong coupling regime, so that the
thermodynamic quantities show universal behavior. This regime occurs
when the Kondo temperature is smaller than the distance of the Fermi
level from the band edge, as quantified by $\mu$.

All other cases, where $\Delta E \propto T_K$, are an indication of
unconvential local moment quenching mechanisms, either through the
trivial local singlet formation for large-$J$, or through an
unconventional flow of the RG equations due to extreme particle-hole
asymmetry due to diverging DOS, as in the 1D case. To conclude: as
soon as the physics is appeciably and simultaneously affected by the
band on all energy scales, including the states at the very edge of
the band, so that the energy scales can no longer be nicely separated
into logarithmic chunks, the binding energy $\Delta E$ and the local
moment quenching scale $T_K$ become equivalent.

\section{Kondo temperature at low density}

\subsection{Prefactors}

\begin{figure}
\centering
\includegraphics[clip,width=0.5\textwidth]{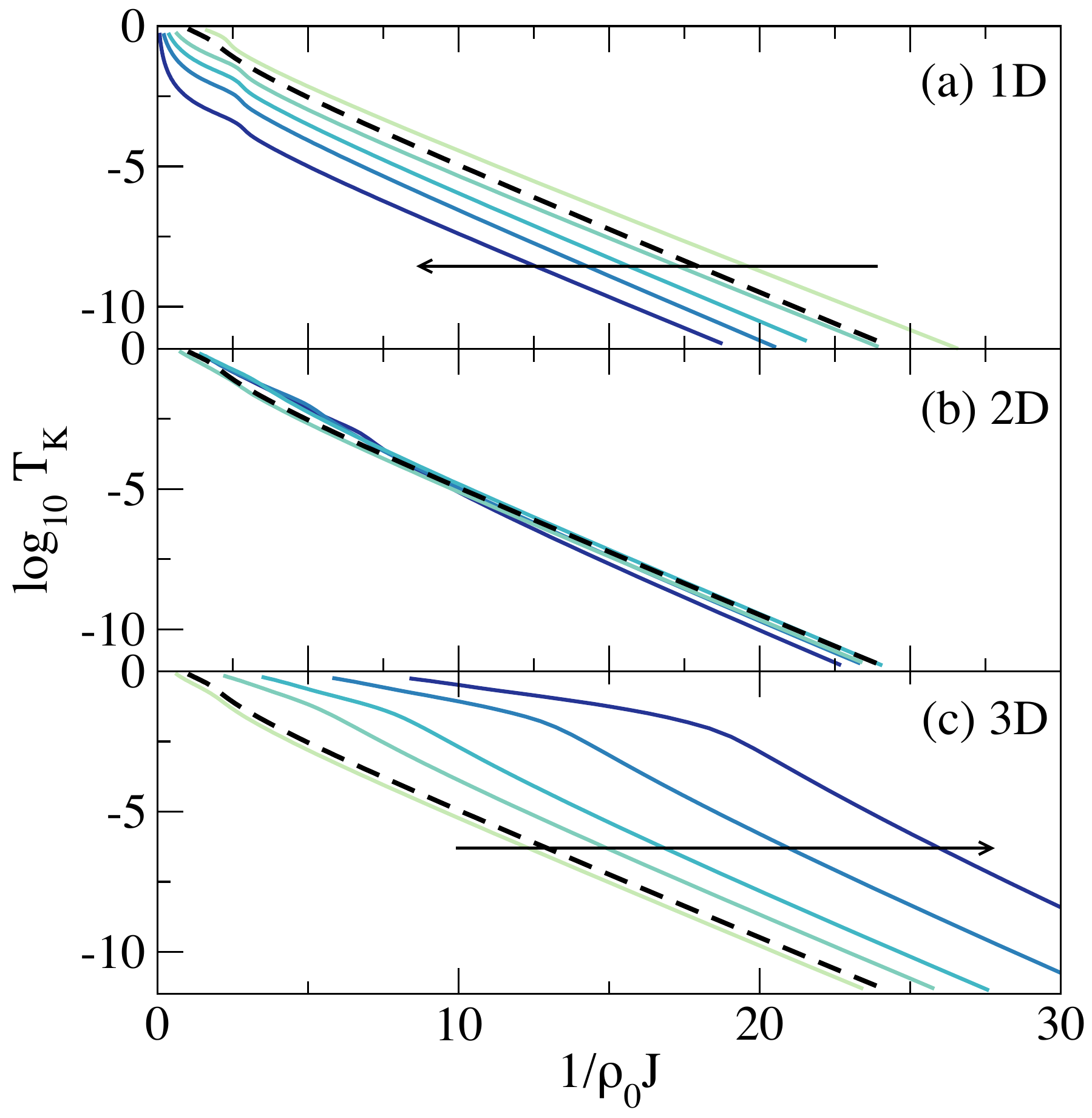}
\caption{Asymptotic scaling of the Kondo temperature. We plot the
logarithm of $T_K$ versus $1/\rho_0 J$, where $\rho_0$ is the density
of states in the conduction band at the Fermi level,
$\rho_0=\rho(\mu)$. The dashed line indicates the standard result for
a flat density of states. The arrows indicate the direciton of reduced
band filling $n$, i.e., going toward a more asymmetric situation. In
(a) we plot $n=0.012$, $0.041$, $0.074$, $0.12$, and $0.5$, in (b) we
plot $n=0.0011$, $0.0083$, $0.085$, and $0.39$, (c) we plot
$n=0.0013$, $0.0032$, $0.012$, $0.035$, and $0.5$.}
\label{fig3}
\end{figure}

Finally, we discuss the quantitative effects of the band-edge
singularities in the DOS on the scaling of the Kondo temperature
against $J$. We plot the logarithm of the Kondo temperature versus
$1/\rho_0 J$, where $\rho_0$ is the density of states at the Fermi
level, for a range of $n$ from the half-filling, $n=1/2$, to very
small values of the density, see Fig.~\ref{fig3}. The non-universal
behavior for intermediate and large $J$ is due to band-edge
singularities and therefore depends on the dimensionality of the
system. For small $J$, however, the exponential behavior $T_K \sim
\exp(-1/\rho_0 J)$ is always eventually recovered and we find straight
lines with essentially equal slope. We emphasize that the main effect
of the $n$-dependence is already included through the DOS at the Fermi
level, $\rho_0=\rho(\mu)$, thus the trend indicated by the arrows in
the figure (oriented from half-filling to small $n$ range) is due to
the differences contained in the prefactor $c_d(\mu)$ to the exponential
term in the expression for $T_K$:
\begin{equation*}
T_K = c_d(\mu) \exp[-1/\rho_0 J].
\end{equation*}
Here we find notable differences that can be ascribed solely to the
dimensionality-dependent singularities in the DOS: in 1D, the
prefactor $c_1$ at constant $\rho_0 J$ decreases with decreasing
filling $n$, leading to lower $T_K$, while exactly the opposite
behavior is found in the prefactor $c_3$ for the 3D case; the 2D case
is in the intermediate situation with curves that are overlapping to a
good approximation, hence $c_2$ is approximately constant. To be more
precise, close to the band-edge (for $\mu$ small compared to the
half-bandwidth $D=2dt$), we find that the prefactors $c_d(\mu)$ can be
approximated fairly well as
\begin{equation}
\label{ls}
\begin{split}
c_1 &\approx \mu, \\
c_2 &\approx D, \\
c_3 &\approx 25.5 D \exp[-9.5 \sqrt{\mu/D}].
\end{split}
\end{equation}

\subsection{Scaling equations for particle-hole asymmetric band}

The differences between the three dimensionalities are too large to be
ascribed solely to the effect of the filling-dependence of the
effective bandwidth. Instead, they must be ascribed to the
renormalization of the exchange coupling due to potential scattering,
$\rho_0 J \to \rho_0 J_\mathrm{eff}$. There is no bare potential term
in the Hamiltonian, but it is generated by the renormalization flow
when the particle-hole symmetry is broken away from half-filling. We
now provide an analytical account of this behavior. The scaling
equations for the spin-$1/2$ Kondo model with the Hamiltonian
expressed in the form
\begin{equation}
H_\mathrm{imp}=J \vc{S} \cdot \vc{s} + V n,
\end{equation}
where $V$ is the potential and $n=\sum_a f^\dag_{a} f_a$ is the local
density of bulk electrons at the position of the impurity, are
\cite{hewson,Ujsaghy:2002by}
\begin{equation}
\begin{split}
\frac{dJ}{d\ln D} &= - \frac{\rho_+}{2} J^2 + 2\rho_- JV, \\
\frac{dV}{d\ln D} &= \rho_- \left( \frac{3}{16} J^2 + V^2 \right),
\end{split}
\end{equation}
where
\begin{equation}
\begin{split}
\rho_+ &= \rho(D)+\rho(-D), \\
\rho_- &= \rho(D)-\rho(-D),
\end{split}
\end{equation}
are the symmetric and antisymmetric combination of the density of
states in the empty and filled parts of the band, with the energy
argument $\epsilon$ of $\rho(\epsilon)$ now measured with respect to
the Fermi level. These equations are fully general and are valid for
arbitrary $\rho(\epsilon)$.

It has been shown that the potential scattering in asymmetric bands
with non-zero $\rho_-$ may play an important role
\cite{Ujsaghy:2002by}. By inspection of the scaling equation for $J$
we see that if $V$ and $\rho_-$ are of the same sign, the growth of
$J$ with decreasing bandwidth slows down ($T_K$ is reduced), and the
opposite is the case if $V$ and $\rho_-$ are of opposite signs ($T_K$
is increased). The scaling equation for $V$ tells that if the bare $V$
is zero, effective potential scattering will be generated by the
exchange scattering in second order so that $V$ is of the {\it
opposite} sign as $\rho_-$. Thus, an asymmetric DOS might be expected
to always lead to a higher Kondo temperature irrespective of the sign
of $\rho_-$.

\begin{figure}
\centering
\includegraphics[clip,width=0.5\textwidth]{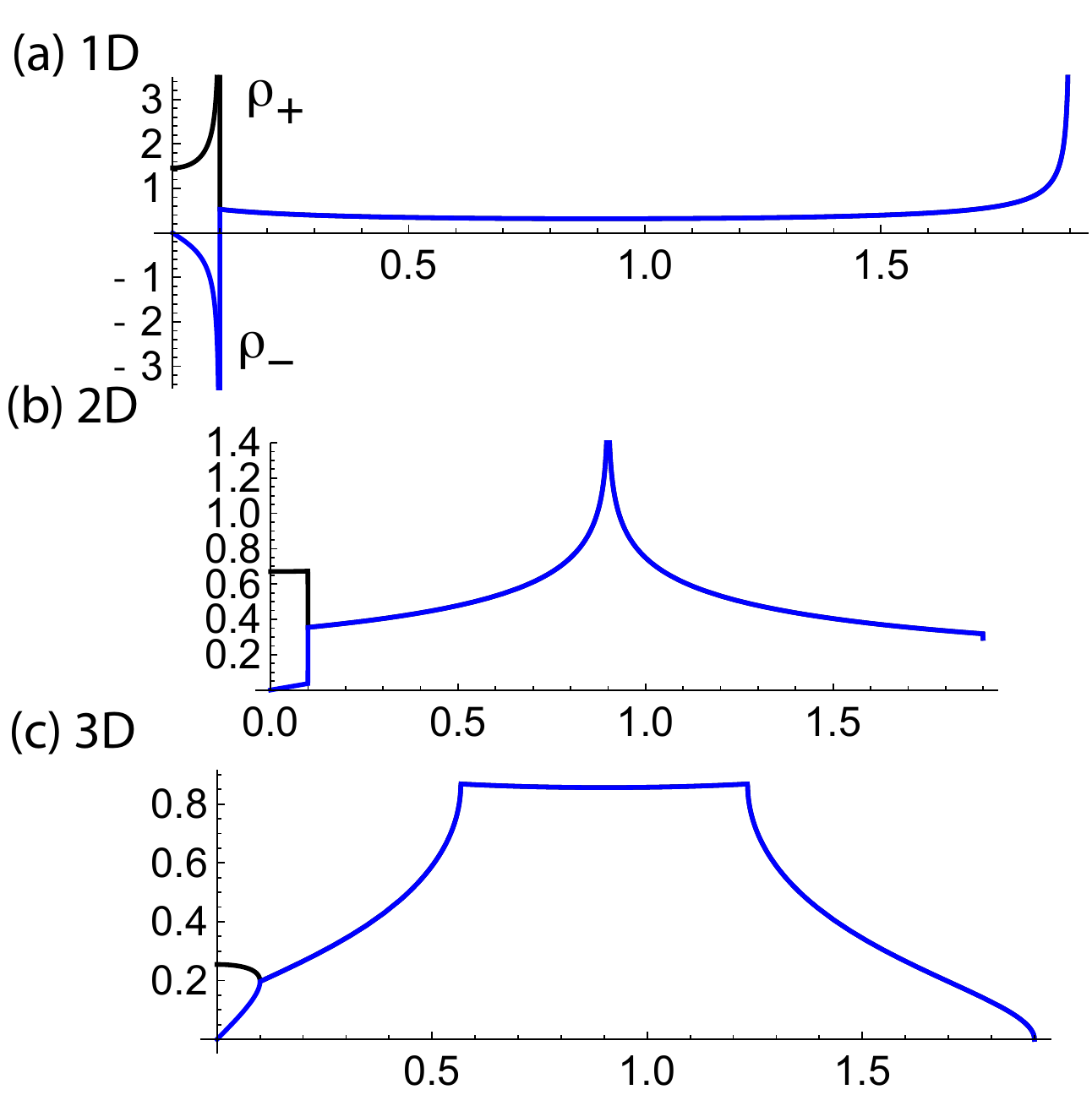}
\caption{Symmetric and antisymmetric combination of the
conduction-band density of states that affects the renormalization
flow: $\rho_\pm(\epsilon)=\rho(\epsilon)\pm\rho(-\epsilon)$. Here
$\mu=0.1$.}
\label{dos}
\end{figure}

We now consider our actual problem to see that the situation is
slightly more subtle. There are two energy regions. In the high-energy
region (i) for $|\epsilon|>\mu$ only the renormalization due to the
non-occupied high-energy states between $\mu$ and the upper band edge
$D$ contributes. Then $\rho_+(\epsilon)=\rho_-(\epsilon) =
\rho(+\epsilon)>0$, see Fig.~\ref{dos}. This will generate a potential
scattering $V$ with a negative sign. While the detailed form of $\rho$
depends on the dimensionality, the qualitative behavior is the same in
all three cases.

In the low-energy region (ii) for $|\epsilon|<\mu$ there will be both
particle-like and hole-like processes. The behavior of $\rho_+$ and
$\rho_-$ in this region strongly depends on the dimensionality,
because close to the band edge $\rho$ is concave in 3D, approximately
flat in 2D, and convex in 3D. For this reason, in 3D $\rho_-$ is
positive, in 2D $\rho_-$ is approximately zero, while in 1D $\rho_-$
is negative, see Fig.~\ref{dos}, and these differences become
increasingly pronounced the closer $\mu$ gets to the band edge due to
the curvature of the DOS. Thus, in 3D $V \rho_- < 0$ and $T_K$ will
tend to be increased, as indeed observed in Fig.~\ref{fig3}(c) for
increasingly asymmetric band (smaller electron density), as indicated
by the arrow. In 1D, however, $V \rho_- > 0$ and $T_K$ will be
reduced, again in line with the NRG results in Fig.~\ref{fig3}(a).
Finally, in 2D with $\rho_- \approx 0$ due to the approximate
flatness, the potential scattering term does not play a significant
role in the renormalization of $J$, thus we recover results which
nearly overlap with those obtained in the flat-band approximation, see
Fig.~\ref{fig3}(b).

The numerical solutions of the scaling equations can indeed be fitted
to $T_K = c_d(\mu) \exp(-1/\rho_0 J)$ with $c_d(\mu)$ given by
Eqs.~\eqref{ls} with some small deviations due to the scaling
equations being truncated at the second order in $J$ and $V$, while
the NRG includes processes to all orders.

\section{Conclusion}

This work explored two issues: 1) the characteristic low-energy scales
of the Kondo impurity model, focusing in particular on the difference
between the binding energy $\Delta E$ and the local-moment quenching
scale $T_K$, 2) the effects due to the band-edge van Hove
singularities characteristic of the different dimensionalities of the
conduction band, which lead to significant effects in the regime of
very low electron density. We showed that in the asymptotic
weak-coupling low-$J$ (scaling) regime the binding energy is quadratic
in $J$, while the Kondo temperature is exponentially small. While it
is meaningful to compare the scale $T_K$ to other magnetic coupling
scales (such as the RKKY coupling $J_\mathrm{RKKY}$) when discussing
the competition between the Kondo screening and magnetic ordering,
because this decides the fate of the effective moment for temperatures
below $\mathrm{max}(T_K, J_\mathrm{RKKY})$, this does not imply that
an isolated impurity reduces the total ground state energy only by
$T_K$. Instead, $T_K$ is only a minor correction to the total energy
gain arising from the exchange coupling of the impurity with bulk
electrons, which is $\propto J^2$. As concerns the dimensionality, we
find that the case of 2D is the closest to the conventional Kondo
scenario, because in the relevant low-energy range the 2D DOS is
approximately flat. For low $J$, the Kondo temperature is thus given
by the standard expression $T_K \approx D \exp(-1/\rho_0 J)$. In 1D
and 3D we find notable deviation in opposing directions. For the 3D
case with concave DOS, the Kondo temperature is increased for reduced
band filling: $T_K \approx 25.5 D \exp[-9.5 \sqrt{\mu/D}]
\exp(-1/\rho_0 J)$. For the 1D case, we find that the Kondo
temperature in some range of exchange couplings $J$ such that $T_K >
\mu$ is a quadratic function of $J$, while for small $J$ the
exponential dependence is recovered. Since the 1D DOS is convex, we
find that the Kondo temperature decreases compared to the standard
flat-band value for reduced band filling: $T_K \approx \mu
\exp(-1/\rho_0 J)$.

Similar trends are expected for other impurity models such as the
Anderson impurity model. There the detailed behavior will also depend
on the intrinsic potential scattering (bare $V$ after the
Schrieffer-Wolff transformation). It should also be noted that in the
context of the dynamical mean-field theory (DMFT) the effective
impurity model is actually in the intermediate coupling regime where
$\Delta E \propto T_K$, thus the energy gain and the coherence
temperature are expected to be of the same scale.

\begin{acknowledgments}
The authors acknowledge the support of the Slovenian Research Agency
(ARRS) under P1-0044 and J1-7259, and thank Jernej Mravlje for
comments.
\end{acknowledgments}

\appendix

\section{Magnetic field effects}

\begin{figure}
\centering
\includegraphics[clip,width=0.5\textwidth]{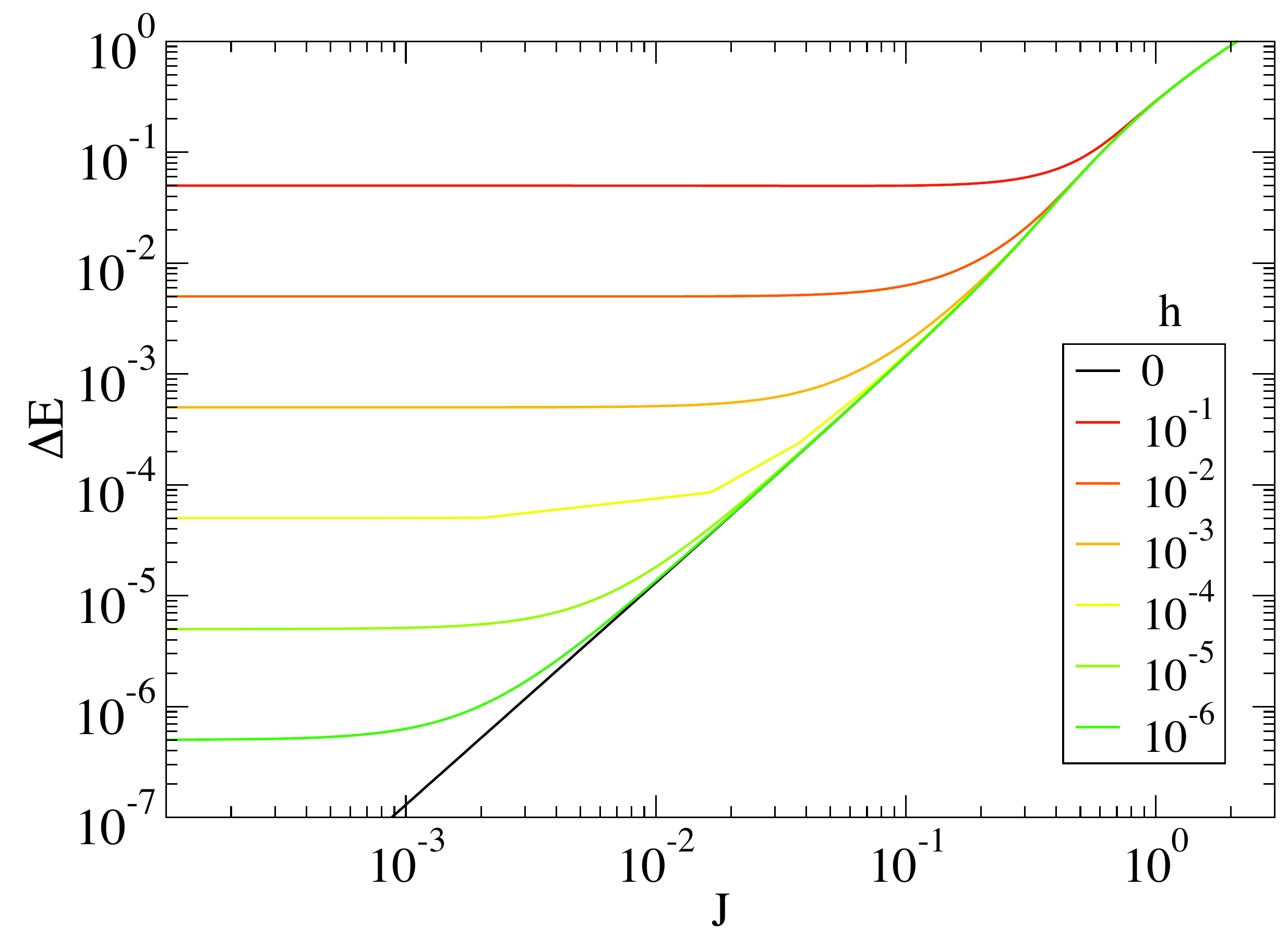}
\caption{Energy gain $\Delta E$ in the presence of the magnetic field.
Here we use a flat band and the chemical potential is fixed in the
center of the band.}
\label{field}
\end{figure}

We now briefly consider the energy gain of the impurity in the
presence of the Zeeman term in the Hamiltonian:
\begin{equation}
H_{\mathrm{Zeeman}} = g_\mathrm{imp} \mu_B B S_z
+ \sum_{k} g_\mathrm{bulk} \mu_B B s_{z,k}.
\end{equation}
We take $g \equiv g_\mathrm{imp}=g_\mathrm{bulk}$, and express the
field in units of the Zeeman energy, $h=g\mu_B B$. The results,
displayed in Fig.~\ref{field}, show the expected result: for small $J$
the energy gain saturates at the value $\Delta E=h/2$ with the
cross-over occuring for $J$ such that $\Delta E \approx h/2$. The
value of $T_K$ does not play any role here.

\section{Anisotropic Kondo coupling}

\begin{figure}
\centering
\includegraphics[clip,width=0.5\textwidth]{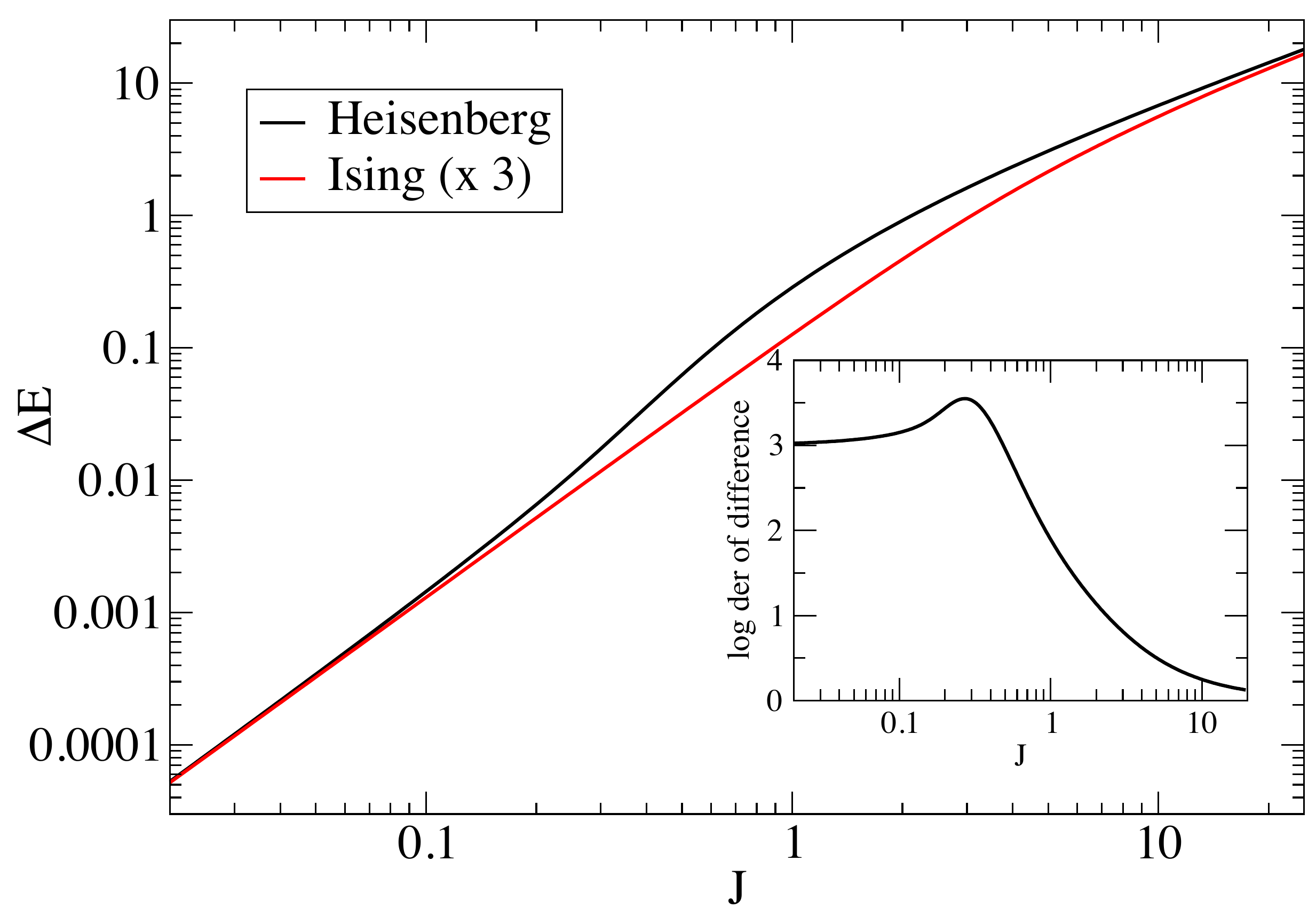}
\caption{Energy gain $\Delta E$ in anisotropic Kondo model with
exchange coupling of Ising type, $J_\|=J$ and $J_\perp=0$, and in the
isotropic model with exchange coupling of Heisenberg type,
$J_\|=J_\perp=J$. The result for the Ising coupling is multiplied by 3
for easier comparison. We use a flat band with $\mu$ fixed in its
center. The inset shows the logarithmic derivative of $(\Delta
E)_\mathrm{Heisenberg}-3(\Delta E)_\mathrm{Ising}$, thus indicating
the power-law exponent of the correction due to spin-flip dynamics. }
\label{ising}
\end{figure}

The XXZ anisotropic Kondo model takes the following form:
\begin{equation}
H_{\mathrm{XXZ}} = J_\| S_z s_z + J_\perp (S_x s_x + S_y s_y).
\end{equation}
The extreme case is that of Ising-like coupling, $J_\perp \equiv 0$.
There is no impurity dynamics in this limit and the Hamiltonian
becomes quadratic, hence exactly diagonalisable (see the following
Appendix). In the small-$J$ and high-$J$ limits, the energy gain for
Ising coupling with $J_\|=J$ is one third of that in the regular
isotropic Kondo model with Heisenberg coupling $J_\|=J_\perp=J$, while
the non-trivial deviations due to many-particle physics occur in the
intermediate-$J$ range, see Fig.~\ref{ising}. The energy gain is
always larger in the isotropic model even after accounting for the
overall factor of 3. In the small-$J$ asymptotic regime where $\Delta 
E_\mathrm{Heisenberg} \approx 3\times\Delta E_\mathrm{Ising}$ to a good
approximation, the Kondo temperature is smaller than $\Delta E$
already by many orders of the magnitude.

We have also computed the difference of the energy gain for isotropic
exchange and three times the energy gain for Ising coupling. The
result is plotted as the inset in Fig.~\ref{ising} in the form of the
logarithmic derivative of the quantity. Considering that the quadratic
term in the energy gain is entirely due to the spin-dependent
scattering on a static local moment, we see that the leading
contribution to the ground state energy of the spin-flip processes
leading to the Kondo effect is of the order of $J^3$ (and not $T_K$). 

\section{Energy gain in the $J\to0$ limit}

The energy gain in the small-$J$ limit will now be calculated
analytically. The calculation is performed using the equations of
motion for a static impurity in the bulk:
\begin{equation}
\begin{split}
H_\mathrm{kin} &= \sum_{k\sigma} \epsilon_k c^\dag_{k\sigma}
c_{k\sigma},\\
H_\mathrm{int} &=(2h) s_z(\vc{r}=0),
\end{split}
\end{equation}
The factor 2 in $(2h)$ is included for convenience. For $S=1/2$
impurity, $h=J/4$. We now drop the spin index $\sigma$ and focus on
the $\sigma=\uparrow$ case; the sums over spin are performed by taking
$h \to -h$ for $\sigma=\downarrow$. The energy gain is defined as
\begin{equation}
\Delta E = \expv{H}_{J=0} - \expv{H}_{J},
\end{equation}
and can be split into two contributions:
\begin{equation}
\Delta E = (\Delta E)_\mathrm{kin} + (\Delta E)_\mathrm{int}.
\end{equation}
For the impurity $T$-matrix we find
\begin{equation}
T(z) =
\frac{1}{N}
\left[
h + h^2 \corr{f;f^\dag}_z
\right],
\end{equation}
where $N$ is the number of lattice sites in the system, and
$f=(1/\sqrt{N})\sum_k c_k$, so that the $k$-resolved Green's function
is
\begin{equation}
G_{kk'}(z) = G^0_{k}(z) \delta_{kk'}
+ G^0_k(z) \frac{1}{N} \left[
h + h^2 \corr{f;f^\dag}_z
\right] G^0_{k'}(z),
\end{equation}
with $G^0_{k}(z)=(z-\epsilon_k)^{-1}$ being the non-perturbed
bulk Green's function. We note that
\begin{equation}
G_f(z) \equiv \corr{f;f^\dag}_z = \frac{1}{N} \sum_{kk'} G_{kk'}(z),
\end{equation}
hence we sum over $k$ and $k'$ to obtain
\begin{equation}
G_f(z) = G^0_\mathrm{loc}(z) + G^0_\mathrm{loc}(z)
\left[ h + h^2 G_f(z) \right] G^0_\mathrm{loc}(z).
\end{equation}
The solution is
\begin{equation}
G_f(z) = \frac{G^0_\mathrm{loc}(z)}{1-h G^0_\mathrm{loc}(z)}.
\end{equation}
For a flat band with half-bandwidth $D$, the exact expression for
$G^0_\mathrm{loc}(z)$ is
\begin{equation}
G^0_\mathrm{loc}(z) = -\frac{1}{2D} \ln \frac{z-D}{z+D}.
\end{equation}
We now calculate the energy gain due to the interaction term,
$(\Delta E)_\mathrm{int} = - \expv{ H_\mathrm{int} }
= -h ( n_{f\uparrow} - n_{f\downarrow} )$,
with
\begin{equation}
n_{f\sigma} = \expv{f^\dag_\sigma f_\sigma} =
\int_{-\infty}^{\infty} f(\omega)d\omega\,
\frac{-1}{\pi} \Im G_{f\sigma}(z),
\end{equation}
$f(\omega)$ being the Fermi function.
We take the difference
\begin{equation}
h \left[ G_{f\uparrow}(z) + (h\to-h) \right] = \frac{2 h^2
G^0_\mathrm{loc}(z)}{1-h^2
[G^0_\mathrm{loc}(z)]^2} \approx 2 h^2 [G^0_\mathrm{loc}(z)]^2.
\end{equation}
Then
\begin{equation}
(\Delta E)_\mathrm{int} = \frac{2h^2}{\pi} \int_{-\infty}^{\infty} \Im 
\left\{
\left[
G^0_\mathrm{loc}(\omega+i\delta) \right]^2 \right\}\, f(\omega)\,d\omega.
\end{equation}
Noting that for $z=\omega+i\delta$,
\begin{equation}
\Im [G^0_\mathrm{loc}(z)]^2 = 2 \Im
G^0_\mathrm{loc}(z) \Re
G^0_\mathrm{loc}(z) = 2 \pi \rho^2 \ln\frac{D-\omega}{D+\omega},
\end{equation}
and integrating over $\omega$ at $T=0$, we finally find
\begin{equation}
(\Delta E)_\mathrm{int} = 8\ln2 (\rho h)^2 = \frac{\ln 2}{8} (J/D)^2.
\end{equation}
The total energy gain can now be computed using the formula
\begin{equation}
\Delta E = \int_0^J \frac{dJ'}{J'} \expv{ H_1 }_{J'},
\end{equation}
where the expectation value is that of the interaction part of the
Hamiltonian, here equal to $(\Delta E)_\mathrm{int}$ evaluated at
$J=J'$, see also Eq.~\eqref{eqnr} with $\beta=2$. Alternatively, we can
explicitly calculate the band contribution $(\Delta E)_\mathrm{bulk}$.
The kinetic energy is
\begin{equation}
E_{\mathrm{kin}}=\sum_k \epsilon_k n_{k},
\end{equation}
with
\begin{equation}
n_{k} = \int_{-\infty}^{\infty} A_{k}(\omega) f(\omega)
d\omega,
\end{equation}
and
\begin{equation}
A_{k}(\omega) = -\frac{1}{\pi} \Im G_{kk}(\omega+i \delta).
\end{equation}
\begin{widetext}
The first term in $G_{kk}$ cancels out after subtracting the energy of
the system without the impurity.
Thus
\begin{equation}
(\Delta E)_\mathrm{kin,\sigma} = \frac{1}{\pi}  \sum_k \epsilon_k 
\int_{-\infty}^{\infty}
d\omega\,
\Im\left( \frac{1}{z-\epsilon_k} \frac{1}{N} \left[ h+h^2 G_f(z) \right]
\frac{1}{z-\epsilon_k} \right) f(\omega).
\end{equation}
This has to be summed over spin. Recalling that $h$ changes sign, the
first term cancels out and 
\begin{equation}
h^2 \left[ G_f(z) + (h\to-h) \right] = 2h^2
\frac{G^0_\mathrm{loc}(z)}{1-h^2
[G^0_\mathrm{loc}(z)]^2} \approx 2h^2 G^0_\mathrm{loc}(z).
\end{equation}
For small $h$ we are left with ($z=\omega+i\delta$)
\begin{equation}
\begin{split}
\Delta E_1 &= 2\frac{h^2}{\pi} \frac{1}{N} \sum_k \epsilon_k
\int_{-\infty}^{\infty}
d\omega\,
\Im\left( \frac{1}{z-\epsilon_k} G^0_\mathrm{loc}(z) 
\frac{1}{z-\epsilon_k} \right) f(\omega) \\
&= 2\frac{h^2}{\pi} \int \epsilon \rho(\epsilon) d\epsilon\,
\int_{-\infty}^{\infty}
d\omega\,
\Im\left( \frac{1}{z-\epsilon} G^0_\mathrm{loc}(z) 
\frac{1}{z-\epsilon} \right) f(\omega).
\end{split}
\end{equation}
\end{widetext}
The $\epsilon$-integral can be evaluated for a flat band (using $D=1$):
\begin{equation}
\int_{-1}^1 \frac{\epsilon}{(z-\epsilon)^2} d\epsilon =
\frac{2z}{z^2-1}+\ln\frac{z-1}{z+1}
\equiv F(z).
\end{equation}
Then at $T=0$
\begin{equation}
(\Delta E)_\mathrm{kin} = -\frac{2(h\rho)^2}{\pi} \Im \left( \int_{-\infty}^0
F(\omega+i\delta) g(\omega+i\delta) d\omega \right),
\end{equation}
where $g(z) = -\ln\frac{z-D}{z+D}$. The integration gives $2\pi \ln2$.
Thus 
\begin{equation}
(\Delta E)_\mathrm{kin}=-\frac{\ln 2}{16} (J/D)^2.
\end{equation}

We finally obtain
\begin{equation}
\Delta E = \frac{\ln 2}{16} (J/D)^2.
\end{equation}
This agrees within a few permil with the numerical renormalization
group results for the gain in the ground state energy in the small-$J$
limit of an Ising-coupled magnetic impurity. For a full isotropic
coupling, we find exactly three times as much:
\begin{equation}
\Delta E = \frac{3\ln 2}{16} (J/D)^2,
\end{equation}
again in full agreement with the numerical calculation. The leading
correction due to dynamic processes is $O(J^3)$, as demonstrated in
the previous Appendix.

\bibliography{tk}

\clearpage
\newpage

\end{document}